\documentclass[prl,nofootinbib,superscriptaddress,twocolumn]{revtex4}
\def\mysection#1{{\bf #1.} }
\def\mysections#1{{\bf #1.} }

\usepackage{amssymb}
\usepackage{amsmath}
\usepackage{graphicx}
\usepackage{longtable}
\usepackage{verbatim}
\usepackage{amsfonts}

\newcommand{\be}{\begin{equation}}
\newcommand{\ee}{\end{equation}}
\newcommand{\bea}{\begin{eqnarray}}
\newcommand{\eea}{\end{eqnarray}}
\newcommand{\beq}{\begin{equation}}
\newcommand{\eeq}{\end{equation}}
\def\beqa{\begin{eqnarray}}
\def\eeqa{\end{eqnarray}}
\newcommand{\no}{\nonumber}
\def\lsim{\mathrel{\rlap{\lower4pt\hbox{\hskip1pt$\sim$}}
    \raise1pt\hbox{$<$}}}         
\def\gsim{\mathrel{\rlap{\lower4pt\hbox{\hskip1pt$\sim$}}
    \raise1pt\hbox{$>$}}}         

\begin{document}

\vspace*{-30mm}

\title{\boldmath Asymmetric Higgsino Dark Matter}

\author{Kfir Blum}\email{kblum@ias.edu}
\affiliation{Institute for Advanced Study, Princeton 08540, USA}

\author{Aielet Efrati}\email{aielet.efrati@weizmann.ac.il}
\affiliation{Department of Particle Physics and Astrophysics,
  Weizmann Institute of Science, Rehovot 76100, Israel}

\author{Yuval Grossman}\email{yg73@cornell.edu}
\affiliation{Institute for High Energy Phenomenology, Newman
  Laboratory of Elementary Particle Physics, Cornell University,
  Ithaca, NY 14853, USA}

\author{Yosef Nir}\email{yosef.nir@weizmann.ac.il}
\affiliation{Department of Particle Physics and Astrophysics,
  Weizmann Institute of Science, Rehovot 76100, Israel}

\author{Antonio Riotto}\email{antonio.riotto@unige.ch}
\affiliation{D\'epartement de Physique Th\'eorique and Centre for
  Astroparticle Physics (CAP),\\
24 quai E. Ansermet, CH-1211 Gen\`eve, Suisse}
\vspace*{1cm}

\begin{abstract}
\noindent
In the supersymmetric framework, a higgsino asymmetry exists in the
universe before the electroweak phase transition. We investigate
whether the higgsino is a viable asymmetric dark matter candidate. We
find that this is indeed possible. The gauginos, squarks and sleptons
must all be very heavy, such that the only electroweak-scale
superpartners are the higgsinos. The temperature of the electroweak
phase transition must be in the $(1-10)$ GeV range.
\end{abstract}

\maketitle

\noindent
\mysection{Introduction}
\noindent
The matter content of our universe is made of two main components: the
dark matter (DM), with $\Omega_{\rm DM} \sim 0.20$, and baryons, with
$\Omega_{\rm b}\sim0.04$. Intriguingly, neither of these numbers can
be explained within the Standard Model (SM) of particle physics. The
most intensively studied scenarios involve very different mechanisms
to explain these two numbers. The DM relic abundance is explained by a
freeze-out of a weakly interacting massive particle number density
that occurs when its annihilation rate becomes slower than the
expansion rate of the universe. The baryon relic abundance is
explained by an asymmetry between baryons and anti-baryons. Under
these circumstances, there is no explanation of the fact that the
energy densities of the DM and the baryons are surprisingly close to
each other, which is then just a coincidence. It would be more
satisfying if it could be naturally explained. This can be the case if
the DM density were also the result of an asymmetry, rather than of
freeze-out
\cite{Nussinov:1985xr,Kaplan:1991ah,Kaplan:2009ag,Farrar:2005zd,Hooper:2004dc,Kitano:2004sv,Kitano:2008tk,Chang:2009sv,Kribs:2009fy,An:2009vq,Cohen:2010kn,Falkowski:2011xh}.
This type of scenarios comes under the name of asymmetric dark matter.

One of the best motivated extensions of the SM is the Minimal
Supersymmetric Standard Model (MSSM). Extending this framework, there
are several ways to generate the baryon asymmetry. Perhaps the most
plausible one is via leptogenesis (for a review, see
\cite{Davidson:2008bu}). Prior to the ElectroWeak Phase Transition
(EWPT), when the higgsino is of Dirac nature, leptogenesis generates
also Higgs and higgsino asymmetries that are initially of similar size to the
lepton asymmetry (see {\it e.g.}  \cite{Nardi:2005hs}).  Regardless of
the source of the baryon and higgsino asymmetries, however, the
conditions of chemical equilibrium imply that, at temperatures above
the EWPT, a non-zero baryon asymmetry requires that there is also a
non-zero higgsino asymmetry.  Could such a higgsino asymmetry survive
and lead to asymmetric higgsino dark matter after the EWPT, when the
higgsino becomes of Majorana nature? This is the question that we
address in this work.

Imagine that the following set of conditions applies:
1) the higgsino is the lightest supersymmetric particle;
2) before the EWPT, higgsino number-changing interactions become
  slow enough that the higgsino asymmetry remains constant;
3) after the EWPT, higgsino-antihiggsino oscillations are slow or,
  if they are fast, the higgsino-antihiggsino annihilation rate is
  slow.
Then, a rather large relic higgsino density can survive.  In this
work, we study the constraints on the supersymmetric spectrum and on
cosmological parameters that follow from imposing this set of
conditions.

\noindent
\mysection{The higgsino asymmetry}
The asymmetry in a particle $x$ number density, $\Delta n_x\equiv
(n_x-n_{\bar x})$, is related to its chemical potential, $\mu_x$, via
(for $\mu_x\ll T$)
\beq\label{eq:asycp}
\Delta n_x^{\rm eq} = \frac{ g_x T^2 \mu_x} {6} K(z_x),
\eeq
where $g_x$ is the number of internal degrees of freedom of the
particle $x$, $z_x\equiv m_x/T$, $K(z_x\ll 1)=2\,(1)$ for bosons
(fermions) and $K(z_x\gg 1)$ is exponentially suppressed.  We are
interested in relating the higgsino asymmetry to the baryon asymmetry.
Under the conditions that will be of interest to us (see below), all
of the sfermions are much heavier than the higgsinos and the sfermion
number densities are negligible because the relevant $K(z)$ factors
are exponentially suppressed. On the other hand, for the quarks and
leptons $K(z)=1$. With these assumptions, the baryon and lepton
asymmetries are given by
\be\label{eq:defbl}
\Delta Y_B=\frac{T^2}{6}(2\mu_Q+\mu_u+\mu_d),\,
\Delta Y_L=\frac{T^2}{6}(2\mu_L+\mu_e),
\ee
where $\mu_\psi\equiv\sum_{i}\mu_{\psi_i}$ ($i=1,2,3$ is a generation
index).  We define the comoving asymmetry via $ \Delta Y_x \equiv
\Delta n_x /s$, where $s$ is the entropy density. We follow the
derivation of Refs.  \cite{Harvey:1990qw,Chung:2008gv}.  Imposing the
conditions of fast gauge, Yukawa, and sphaleron interactions, the
Majorana nature of the gauginos, the Dirac nature of the higgsinos
(prior to the EWPT), and hypercharge neutrality, we obtain
\beq\label{eq:Yhse}
\frac{ \Delta Y_{\tilde h}}{ \Delta Y_B}=-\frac{2K(\tilde
  h)}{12+3[K(h_u)+K(h_d)+2K(\tilde h)]}.
\eeq
Eq.~(\ref{eq:Yhse}) relies on super-equilibrium (SE), that is chemical
equilibrium between Higgs particles and higgsinos. At the end of the
epoch of SE, a higgsino asymmetry is conserved until the EWPT.

As long as the higgsinos are relativistic, the right hand side of
Eq.~(\ref{eq:Yhse}) is (in absolute value) in the range $(0.10-0.15)$.
If the higgsinos become non-relativistic while SE persists, then
$\Delta Y_{\tilde h}$ is quenched by the $K(\tilde h)$ factor in
Eq.~(\ref{eq:Yhse}). Such quenching of $\Delta Y_{\tilde h}$ is not
allowed if higgsinos are to provide the DM. Thus, we are led to impose
that SE must be broken while higgsinos are still relativistic,
freezing $\Delta Y_{\tilde h}$ at a value
\be\label{eq:dyh}
\Delta Y_{\tilde h}\approx-10^{-11}.
\ee
In the next section we find the conditions on the supersymmetric
spectrum that would lead to early breakdown of SE.

\noindent\mysection{Non-Super-Equilibrium (NSE)}
Our goal is to find the conditions on the particle spectrum such that
NSE occurs while the higgsinos are relativistic, $T_{\rm NSE}>\mu$. We
use $\Gamma<H$ as the rough criterion for an interaction to be out of
equilibrium, where the Hubble expansion rate is given by $H\approx
10\,T^2/m_{\rm Pl}$ and the interaction rate is given by
$\Gamma=n\langle\sigma v\rangle$, $n$ being the number density of
target particles.

Higgsino number is violated in Higgs-Higgs scattering ($hh\to\tilde
h\tilde h$) and Higgs-antihiggsino scattering ($h\tilde h^c\to
h^c\tilde h$). These processes arise from the effective Lagrangian
\beq\label{Leff}
-\mathcal{L}_{\rm eff}=\frac{1}{\Lambda_u}\tilde h_u\tilde
h_uh_u^*h_u^*
+\frac{1}{\Lambda_d}\tilde h_d\tilde h_dh_d^*h_d^*\,,
\eeq
generated at tree level by gaugino and at one loop by quark-squark
diagrams.  Requiring that $h_uh_u\to\tilde h_u\tilde h_u$ is not in
equilibrium at $T\sim\mu$ puts a lower bound on $\Lambda_u$,
\beq\label{eq:susyspectrumL}
\Lambda_u\gsim 3 \times 10^9\ {\rm GeV}
\left(\frac{\mu}{1\,\rm TeV}\right)^{1/2}.
\eeq
If $h_d$ is not heavy and decoupled, then a similar bound holds for
$\Lambda_d$.

The gaugino contributions to~(\ref{Leff}) are given by
\beq
\frac{1}{\Lambda_u}=\frac{1}{\Lambda_d}=\frac{g'^2}{8M_1}+\frac{g^2}{8M_2}.
\eeq
The stop and sbottom contributions are given by
\bea\label{Dstsb}
\frac{1}{\Lambda_u}&=&\frac{3\,\alpha_W^2\,m_t^3}{2\,m_W^4\,s_\beta^4}\,
\sin2\theta_t\,\ln\frac{m_{\tilde t_2}^2}{m_{\tilde
    t_1}^2}\,,\nonumber\\
\frac{1}{\Lambda_d}&=&\frac{3\,\alpha_W^2\,m_b^3}{2\,m_W^4\,c_\beta^4}\,
\sin2\theta_b\,\ln\frac{m_{\tilde b_2}^2}{m_{\tilde b_1}^2}\,.
\eea
Eq.~(\ref{eq:susyspectrumL}), therefore, implies
\beqa\label{eq:susyspectrum}
M_i&\gsim&10^8\ {\rm GeV}
\left(\frac{\mu}{1\,\rm TeV}\right)^{1/2}\,,\no\\
\sin2\theta_t\,\ln\frac{m_{\tilde t_2}^2}{m_{\tilde
    t_1}^2}&\lsim&10^{-6}\left(\frac{1\,\rm TeV}{\mu}\right)^{1/2}\,.
\eeqa

Finally, decoupling the decays and inverse decays of sfermions to
higgsino-fermion~\cite{Chung:2008gv} at $T_{\rm NSE}>\mu$ requires all
of the sfermions to be heavy, $m_{\tilde f}\gsim(10-40)\,\mu$. Here,
the stronger bound refers to top squarks and the weaker bound refers
to the electron super-partners.

We note that since, as shown in the next section, $\mu > T_{\rm
  EWPT}$, the above spectrum guarantees that SE ends before the EWPT.

\noindent
\mysection{The spectrum}
The higgsino spectrum includes two neutral mass eigenstates, with mass
splitting $\Delta m_0$~\cite{Giudice:1995qk}, and a
charged higgsino, split by $\Delta m_+$ from the lightest neutral
higgsino~\cite{Drees:1996pk,Cheung:2005pv}.  As concerns $\Delta m_+$,
this quantity does not violate higgsino number and in our scenario it
is therefore dominated by gauge loops, $\Delta
m_+\approx\frac12\alpha_W s_W^2 m_Z\sim350\ {\rm MeV}$.  The results
of the previous section have, however, interesting implications for
$\Delta m_0$. The operators of Eq. (\ref{Leff}) lead to
\beqa\label{dD}
\Delta m_0&=&\Delta_u+\Delta_d,\\
\Delta_u&=&\langle h_u\rangle^2/{\Lambda_u},\;\;
\Delta_d=\langle h_d\rangle^2/{\Lambda_d}.\no
\eeqa
Eq.~(\ref{eq:susyspectrumL}) implies $\Delta_u\lsim10$ keV. An
analogous bound, $\Delta_d\lsim(10\,{\rm keV})/\tan^2\beta$, applies
unless the second Higgs doublet, $h_d$, is much heavier than the
higgsinos. We now show that experimental constraints lead us to choose
the second possibility, namely $m_{h_d}\gg\mu$.

Constraints on the higgsino spectrum in our scenario come from direct
and indirect DM searches. The dominant higgsino-nucleus interactions
are spin-independent inelastic interactions (see, {\it e.g.},
\cite{Cheung:2005pv,Hisano:2004pv,Beylin:2008zz}). The cross section
with a single nucleon, in the limit of zero mass splitting, is
$\sigma_n\approx10^{-38}\ (10^{-36})\ {\rm cm}^2$ for $\mu=100\
(1000)$ GeV. This is orders of magnitude above current bounds. The
only way to evade these bounds (and maintain higgsino LSP) is by a
large enough mass splitting that will make the inelastic scattering
kinematically forbidden
\cite{TuckerSmith:2001hy,Akimov:2010vk,Farina:2011bh}.  The minimum
mass splitting required is a function of the higgsino mass, but in the
entire range of interest for $\mu$ it is smaller than 400 keV. We
learn that the sbottom contribution to $\Delta m_0$ must be
substantial, $\Delta_d\gsim400\,\rm keV$.  As a result, the
only configuration consistent with early breakdown of SE, $T_{\rm
  NSE}>\mu$, is one with a very heavy second doublet, $m_{h_d}\gg\mu$.

As can be seen from Eqs.~(\ref{Dstsb}) and~(\ref{dD}), requiring
\beq
\Delta m_0\gsim {1\,\rm MeV}
\eeq
constrains the sbottom sector to satisfy,
\beq
\sin2\theta_b\gsim 2 \times 10^{-2}\,\left(\frac{\tan\beta}{20}\right)^{-2}\,.
\eeq

Finally, with $\Delta m_0\gg10$ eV, the present higgsino population is
symmetric (see below), and therefore annihilates and may provide
signals in indirect searches for DM. The annihilation cross section
into $WW$ and $ZZ$ pairs is given by \cite{ArkaniHamed:2006mb}
\beq\label{eq:sigind}
\langle\sigma^{\rm ann}v\rangle\approx10^{-26}\ {\rm cm}^3\ {\rm
  sec}^{-1}\ \left({1\ {\rm TeV}}/{\mu}\right)^2.
\eeq
The bound on the cross section from the Fermi-LAT data
\cite{Ackermann:2011wa}, when compared with Eq. (\ref{eq:sigind}),
requires
\be\label{ind}
\mu\gsim190\ {\rm GeV}.
\ee
%

\noindent
\mysection{Oscillations, damping and expansion}
At the EWPT, the Higgs acquires a VEV, and the higgsinos mix with the
gauginos. The resulting propagation eigenstates change from Dirac to
Majorana fermions, and oscillations begin. On the other hand, incoherent
interactions with the plasma continue and damp the oscillations. In
this section, we study the time evolution of the higgsino system
$(\tilde{h}, \bar{\tilde{h}})$ under the simultaneous effects of
oscillations, annihilations, damping and the expansion of the universe.
To do so, we employ the formalism of the density matrix.  This formalism
was originally developed to study neutrinos
\cite{McKellar:1992ja,Stodolsky:1986dx,Harris:1980zi,Enqvist:1990ad}
and we adapt it to our case. Our equations are consistent with those
of Ref. \cite{Cirelli:2011ac} which deals with closely related issues
(see also \cite{Buckley:2011ye}). The quantum rate equations for
$Y_\mu \equiv n_\mu/s$ (where $n_0 \pm n_3$ are the number densities of the
higgsinos and anti-higgsinos, and $n_1 \mp i n_2 $ are the off diagonal
elements of the density matrix) read
\beqa
\frac{\rm d}{{\rm dlog}z}{\bf Y}&=&-\frac{1}{H}\left( \begin{array}{ccc}
    D & V & 0 \\
    V & D & \Delta m_0 \\
    0 & -\Delta m_0 & 0 \end{array}\right){\bf Y},\no\\
\frac{{\rm d}}{{\rm dlog} z}Y_0&=&\frac{s}{H}\langle\sigma^{\rm ann}v\rangle\no\\
&\times&\left[
  2Y^{\rm eq}\bar{Y}^{\rm eq}-\frac12 Y_0^2+\frac12 Y_3^2+G(Y_1^2+Y_2^2)\right],
\nonumber
\eeqa
where $z\equiv\mu/T$.  The damping (or decoherence) factor $D$, and
the effective matter potential $V$, are given by
\be
D=2\sum_f n_f\langle\sigma_{\tilde h+f\to\tilde h+f} v\rangle,
V={8\zeta(3)\alpha_W\eta_B T^3}/{(\pi m_W^2)}.
\nonumber
\ee
While $D$ is proportional to the elastic scattering cross section and
to the total number density of the massless fermions in the plasma,
$V$ is proportional to the elastic scattering amplitude and to the
fermion-antifermion asymmetry. In the limit of large damping, the
effective rate of oscillation is $\Gamma_{\rm osc} \sim (\Delta m_0
)^2 / D$, as obtained in \cite{Falkowski:2011xh}.  The $G$ factor
measures the ratio between the annihilation cross section with and
without including co-annihilations.

The initial conditions, at the EWPT, are the following:
\beq
Y_0=n_{\tilde h}/s,\ \ Y_1=Y_2=0,\ \ Y_3=-10^{-11},
\eeq
where $n_{\tilde h}$ is the solution of the Boltzmann equations (at
the EWPT) with constant asymmetry, and the value of $Y_3= \Delta
Y_{\tilde h}$ is taken from Eq. (\ref{eq:dyh}).

\noindent
\mysection{Asymmetry-assisted higgsino DM}
We aim to solve for $Y_0(\infty)$ which gives the final total number
density in higgsinos, and for $Y_3(\infty)$ which gives the final
higgsino asymmetry. In particular, to provide $\Omega_{\rm
  DM}h^2\approx0.11$, our scenario needs to fit
\be\label{eq:yobs}
Y_0^{\rm obs}\approx7.6\times10^{-13}\left({1\ {\rm
      TeV}}/{\mu}\right).
\ee
We consider three free parameters: $\mu$, $\Delta m_0$ and $T_{\rm
  EWPT}$ and ask whether there is a range of these parameters where
  such a fit is achieved.
If the asymmetry is washed out before the symmetric decoupling
temperature, $T^{\rm sym}_{\rm dec}\approx\mu/25$, then the present DM relic
density is the standard symmetric one, as if an initial asymmetry was
never generated:
\be\label{eq:ysym}
Y_0^{\rm sym}\approx5.9\times10^{-13}\left({\mu}/{1\ {\rm
      TeV}}\right).
\eeq
Comparing Eqs. (\ref{eq:yobs}) and (\ref{eq:ysym}) we learn that for
$\mu\sim1.1$ TeV, higgsinos can account for DM without asymmetry.
Since in the presence of an asymmetry the total number density is
always larger than in the symmetric case, this puts an upper bound of
1.1 TeV on the higgsino mass. Thus, the range of interest is
$ 190\ {\rm GeV}\leq\mu\leq1.1\ {\rm TeV}$.
Within this range, the lighter the higgsino, the larger the asymmetry
that is required in order to satisfy (\ref{eq:yobs}).

\begin{figure}[!t]
\includegraphics[width=0.35\textwidth,height=4cm]{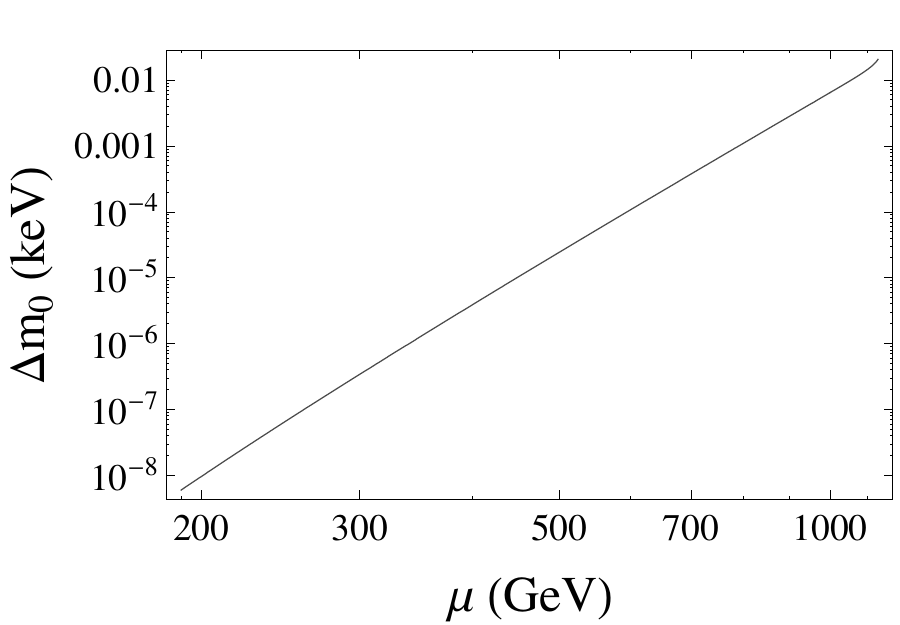}
\caption{The required $\Delta m_0$ as a function
  of $\mu$, with $T_{\rm EWPT}=100$ GeV.}
\label{fig:dmmu}
\end{figure}

We can obtain $Y_0(\infty)>Y_0^{\rm sym}$ if either of the following
two conditions applies:
1) $\Delta m_0$ is small enough that the oscillations are slow, and at
least part of the higgsino asymmetry survives down to $T<T_{\rm
  dec}^{\rm sym}$;
2) The EWPT occurs late enough that annihilations are already slow
when oscillations begin, $T_{\rm EWPT}<T_{\rm dec}^{\rm sym}$.
To study the first possibility, we fix $T_{\rm EWPT}=100$ GeV, and
solve for $Y_0(\infty)$ as a function of $\mu$ and $\Delta m_0$. We
find that the lighter the higgsinos are, the smaller the mass
splitting that is required to provide the DM abundance. The reason is
that smaller $\mu$ leads to smaller $T_{\rm dec}^{\rm sym}$, and the
asymmetry is required to survive to later times. The required $\Delta
m_0$ as a function of $\mu$ is shown in Fig. \ref{fig:dmmu}. It ranges
between $(10^{-2}-10^{-8})$ keV for $\mu$ in the range $(1000-200)$
GeV. Such a small mass splitting is excluded by direct DM searches. We
conclude that it is impossible for the asymmetry to survive once the
EWPT takes place.

\begin{figure}[!t]
\includegraphics[width=0.35\textwidth,height=4cm]{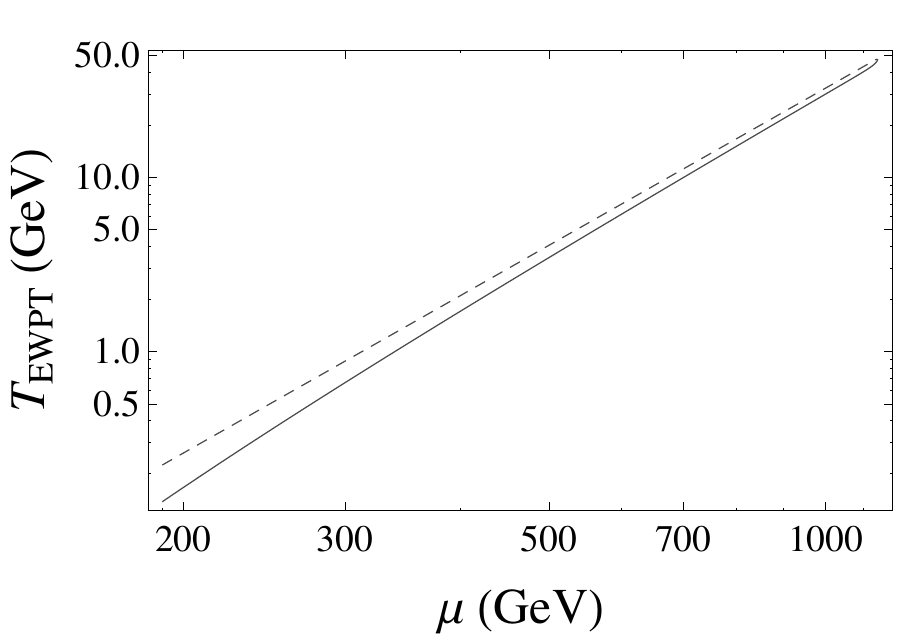}
\caption{The required $T_{\rm EWPT}$ as a function of
  $\mu$, with $\Delta m_0=1$ MeV. The approximate
analytic solution \eqref{eq:tewpt} is shown in dashed line.}
\label{fig:tptmu}
\end{figure}

Second, we fix $\Delta m_0=1$ MeV (the results are not sensitive to
changes in $\Delta m_0$ in the range where it is not excluded by
direct searches), and solve for $Y_0(\infty)$ as a function of $\mu$
and $T_{\rm EWPT}$. Our numerical result for the required $T_{\rm
  EWPT}$ as a function of $\mu$ is shown as the smooth line in
Fig.~\ref{fig:tptmu}. Above the line, the final DM abundance is too
low. Below the line, the final abundance is too high. The required
temperature ranges between $(30-0.1)$ GeV for $\mu$ in the range
$(1000-200)$ GeV.  For $T\gsim3 ~\rm GeV \left( \mu / 1~\rm
TeV\right)^2$, we can obtain an approximate analytic solution for
the required $T_{\rm EWPT}$,
\beq\label{eq:tewpt}
T_{\rm EWPT}=33\ {\rm GeV}\ \left(\frac{\mu}{1\ {\rm TeV}}\right)^3,
\eeq
shown as the dashed line in Fig.~\ref{fig:tptmu}.

We conclude that a viable scenario of asymmetric higgsino dark matter
could have occurred as follows: a higgsino number asymmetry of a size
that is about a factor of 10 smaller than the baryon asymmetry exists
before the EWPT. At the EWPT, the asymmetry is quickly washed out due
to higgsino-antihiggsino oscillations. The resulting symmetric
higgsino population is (for masses below TeV) much larger than the
would-be population without an initial asymmetry. It survives if the
phase transition occurs at a temperature that is somewhat low, of
order (1-10) GeV. We note that such a low temperature requires a rather
strong phase transition, potentially necessitating new degrees of freedom other
than those of the MSSM. Further analysis of this requirement is beyond the scope of the current paper and we postpone it to future work.

\noindent
\mysection{Conclusions}
Within the framework of the MSSM, we ask
whether the higgsino could be a viable asymmetric dark matter
candidate. We find that the answer is in the affirmative, provided
that the following constraints on the supersymmetric spectrum are
satisfied:
\begin{itemize}
\item
Electroweak gauginos are heavier than $10^8$ GeV;
\item
Sfermions are heavier than $10^4$ GeV;
\item
The stop mixing angle is small, and the sbottom mixing angle is large;
\item
Higgsinos are in the range (200-1000) GeV.
\end{itemize}
In addition, the temperature of the electroweak phase transition must
be somewhat low, of order (1-10) GeV.

The supersymmetric spectrum is somewhat reminiscent of split
supersymmetry models~\cite{ArkaniHamed:2004fb,Giudice:2004tc}. The
supersymmetric flavor problem is solved. Grand Unification remains a
viable possibility. Supersymmetry does not solve the fine tuning
problem. It does however explain both the baryon asymmetry and the
dark matter abundance, and relates the two. The initial source of both
asymmetries could be leptogenesis.

This scenario, where the only new particles at the electroweak scale
are the higgsinos, poses a challenge to the LHC. Work on experimental
and observational signals is in progress.

\mysections{Acknowledgments} We thank Nima Arkani-Hamed, Rouven Essig,
Yonit Hochberg, Jesse Thaler and Tomer Volansky for useful
discussions. KB is supported by DOE grant DE-FG02-90ER40542. YG is
supported by NSF grant PHY-0757868 and by a grant from the BSF.  YN is
the Amos de-Shalit chair of theoretical physics and supported by the
Israel Science Foundation (grant \#377/07), and by the German-Israeli
foundation for scientific research and development (GIF).


\end{document}